\documentclass[12pt]{article}

\usepackage{epstopdf}
\usepackage{subfigure}

\usepackage[english]{babel}
\usepackage[ansinew]{inputenc}
\usepackage{epsfig}
\usepackage{color,graphicx}
\usepackage{wrapfig}
\usepackage{amssymb,amsthm,amsmath}
\usepackage{dsfont}
\usepackage{natbib}
\usepackage{abstract}
\usepackage{enumerate}
\usepackage{enumitem}

\renewcommand{\baselinestretch}{1}

\def\ds{\displaystyle}

\begin{document}


{\renewcommand{\baselinestretch}{1}

\title{\bf Bayesian Item Response model: a generalised approach for the abilities' distribution using mixtures}
\author{\bf{Flávio B. Gonçalves$^{a}$, Bárbara C. C. Dias$^{a}$, Tufi M. Soares$^{bc}$}}
\date{}

\maketitle
}

\renewcommand{\abstractname}{Abstract}
\begin{abstract}
Traditional Item Response Theory models assume the distribution of the abilities of the population in study to be Gaussian. However, this may not always be a reasonable assumption, which motivates the development of more general models. This paper presents a generalised approach for the distribution of the abilities in dichotomous 3-parameter Item Response models. A mixture of normal distributions is considered, allowing for features like skewness, multimodality and heavy tails. A solution is proposed to deal with model identifiability issues without compromising the flexibility and practical interpretation of the model. Inference is carried out under the Bayesian Paradigm through a novel MCMC algorithm. The algorithm is designed in a way to favour good mixing and convergence properties and is also suitable for inference in traditional IRT models. The efficiency and applicability of our methodology is illustrated in simulated and real examples.


\end{abstract}

\section{Introduction}\label{section_intro}

Traditional Item Response Theory (IRT) models are based on several assumptions that, although reasonable, are not always met in real applications. Generalisations have been proposed in several directions. They regard the item characteristic curve \citep{bazan2006skew,bazan2010framework,samejima2000logistic}, the abilities' distribution \citep{azevedo2011bayesian}, multilevel structures \citep{goldstein1988general,aitkin1986statistical}, differential item functioning \citep{gonccalves2013simultaneous}, among others.

This paper considers a generalised approach for the abilities' distribution. Traditional IRT models, in which individual and population abilities are to be estimated, consider that distribution to be normal. However, this may not be a reasonable assumption in some cases and more general distributions should be considered, allowing for features like skewness, multimodality and/or heavy tails. For example, the plausibility of the normality assumption is questioned for many psychometric data sets by \citet{micceri1989unicorn}. \citet{schmitt} says that the normality assumption follows if ``one believes that these traits are influenced by a large number of factors, each having a small, equal, and independent effect on the trait." Furthermore, an unreasonable normality assumption may severely corrupt the main analysis \citep[see, for example,][]{seong,kirisci,finch}.

General approaches for the abilities distribution have been previously considered in the literature. For example, \citet{azevedo2011bayesian} assume the abilities to follow a skew normal distribution in the 2 Parameter Normal Ogive model (2PNO). \citet{fox&glas01} propose a multilevel linear regression prior to model the abilities as a function of covariates when the individuals are grouped according to some observed factor. The authors consider the 2PNO model and propose an MCMC algorithm to perform Bayesian inference by using appropriate auxiliary variables that lead to closed form full conditional distributions. Another interesting possible approach is to assume a mixture distribution for the abilities. This has been considered in several works in the context of mixture models, in which the item parameter(s) also vary across latent groups. For example, \citet{bolt&cohen} and \citet{fieuws2004mixture} consider a mixture of normal distributions to model the abilities in the 2PL (with nominal polytomous response) and 1PL models, respectively. The former performs model estimation via MCMC whilst the latter uses Marginal Maximum Likelihood (MML), so only population ability parameters are estimated. Model indeterminacy (non-identifiability) is dealt with through restrictions on the item parameters. \citet{vonDavier23} consider a multidimensional 2PL mixture model with the mixture of normals being a possibility for the abilities' distribution. Estimation is performed via MML and restrictions are imposed to the item parameters to solve the model indeterminacy problem. Some other works propose a non-parametric approach to model the abilities distribution. For example, \citet{schmitt} consider a 2-parameter model and uses standardized Gauss–Hermite quadratures to model the abilities' distribution. \citet{finch} consider the 1PL model and uses a Dihichlet process, with a mixture of normals as an alternative, to model the abilities.

This paper proposes a general approach to model the abilities using a mixture of normal distributions. That is a flexible and analytically simple way to model continuous distributions. Moreover, it allows for relevant practical interpretations in terms of multiple (latent) populations. Unlike previous works, a 3-parameter model - 3PNO, is considered. Inference is performed under the Bayesian Paradigm, which allows for efficient estimation and robust uncertainty quantification of the individual abilities. Model identifiability is achieved through the abilities' distribution, without the need to impose restrictions on the item parameters. A minimum number of restrictions are imposed without compromising the modelling flexibility offered by the mixture of normal distributions. A novel MCMC algorithm is proposed, aiming at improving its convergence properties when compared to existing algorithms.

The efficiency and applicability of our methodology is illustrated in a collection of simulated and real examples. In particular, we consider a data set from the Brazilian High School National Exam (ENEM) which is used for admissions in most of the Brazilian universities. We also consider a data set from PISA regarding the Maths test in 2012 and perform an individual analysis of four countries: Great Britain, South Korea, Poland and United States.

This paper is organised as follows. Section 2 presents our mixture IRT model and Section 3 presents the MCMC algorithm. Results from simulated and real examples are presented in Sections 4 and 5, respectively.

\section{IRT mixture model}\label{section_model}

We consider the 3PNO model for dichotomous items. Particular cases (1 and 2 parameter model) fall automatically under this approach. For a data set with $I$ items and $J$ individuals, let $Y_{ij}$ be the indicator variable of individual $j$ correctly answering item $i$. The model is given by
\begin{equation}\label{m1}
\ds P(Y_{ij}=1|\theta_j,a_i,b_i,c_i)=c_{i}+(1-c_{i})\Phi(a_i\theta_j-b_i),
\end{equation} where $\Phi(.)$ is the standard normal c.d.f. We propose the following mixture distribution for the abilities $\theta_j$:
\begin{equation}\label{2}
\ds (\theta_j | p,\mu, \sigma)\sim\displaystyle \sum_{k=1}^{K}\ p_k N\left(\mu_k,\sigma_{k}^2\right),
\end{equation}
where $N$ denotes the normal distribution, $p=(p_1,\ldots,p_K)$ are the mixture weights, $\mu=(\mu_1,\ldots,\mu_K)$ and $\sigma^2=(\sigma_{1}^2,\ldots,\sigma_{K}^2)$ are the respective means and variances of the mixture components. Define $a=(a_1,\dots,a_I)$ and the analogous notation for $b$ and $c$; also $\theta=(\theta_1,\ldots,\theta_J)$.
The choice for the probit link simplifies the MCMC algorithm as we discuss in Section 3.

Apart from $(\mu_k,\sigma_{k}^2)$, we assume prior independence among all the model parameters, more specifically, $a_i \sim N_{\left(0,\infty \right)}\left(\mu_a,\sigma_a^2\right)$, $b_i \sim N\left(\mu_b,\sigma_b^2\right)$, $c_i\sim Beta(\alpha_c,\beta_c)$, $p \sim Dirichlet(\alpha_1,...,\alpha_K)$ and $(\mu_k,\sigma^2_k) \sim NIG\left(m_k,\frac{1}{\beta},d,e\right)$, where $NIG$ represents the Normal-Inverse Gamma distribution.

Our model is not identifiable due to the non-identifiability of the likelihood function induced by (\ref{m1}). Simply note that any transformation of the type
$\theta^*_j=s(\theta_j+r)$, $a_i=a_i/s$ and $b^*_i=b_i+a_ir$,  $s\in R^+$, $r\in R$, with $j=1,...,J$ and $i=1,...,I$, leads to the same likelihood value for any observed data. Identifying the model can be seen as a way to set its scale and may be achieved by making restrictions on the parameters of the abilities' distribution. For example, in traditional IRT models the mean and variance of the abilities' distribution are usually set to be 0 and 1, respectively (any other values can be chosen). Nevertheless, in the case of our mixture model, fixing all the parameters of the abilities's distribution would seriously compromise the flexibility and applicability of the model. One possible solution would be to fix the mean and variance of the mixture, but this would considerably increase the complexity of the MCMC (or any other estimation method).

We impose some restrictions to the mixture parameters to achieve identifiability without compromising the model's flexibility. Firstly, we fix $(\mu_1,\sigma^2_1)=(0,1)$, which helps to identify the model as long as the first component of the mixture has a reasonably high weight. For that reason, we set $p_1>0.5$. Based on the fact that the traditional models (with only one normal for the abilities) are identified by fixing the parameters of the abilities' distribution, our strategies ought to be enough to resolve the identifiability problems for the proposed mixture model. Basically, the scale of a large group (more the 50\% of the individuals) is fixed by the restriction $(\mu_1,\sigma^2_1)=(0,1)$, therefore identifying the item parameters along that scale which, in turn, identify the abilities of the remaining individuals and, finally, the parameters of possible remaining items. Further label switching problems when $K\geq3$ are avoided by ordering $\mu_2<\mu_3<\ldots<\mu_K$.

Note that the restrictions imposed to the model do not fix its scale \emph{a priori}, like in the traditional model. However, the posterior sample of the abilities may be re-scaled as desired. A re-scaling to mean $m$ and standard deviation $s$ is obtained by making
\begin{equation}
\ds \theta_{j}^{(l)*}=m + s\left(\frac{\theta_{j}^{(l)}-\bar{\mu}^{(l)}}{\bar{\sigma}^{(l)}}\right), \nonumber
\end{equation}
where $(l)$ refers to the $l$-th iteration of the chain, $\bar{\mu}$ and $\bar{\sigma}$ are the mean and standard deviation of the mixture i.e., $\ds\bar{\mu}=\sum_{k=2}^Kp_k\mu_k$ \; and \; $\ds\bar{\sigma}=\sqrt{p_1(\bar{\mu}^2+1)+\sum_{k=2}^Kp_k((\mu_k-\bar{\mu})^2+\sigma_{k}^2)}$.

The number of components $K$ in the mixture may be fixed or estimated. We only consider the former approach in this paper, however, it is straightforward to incorporate any known efficient MCMC algorithm for the estimation of the number of mixture components to our methodology. Moreover, due to the great flexibility of normal mixtures and by what is expected from the population behavior in educational assessment problems, we believe that $K=2$ should be enough to provide a good fit in most cases. Nevertheless, if really required, $K=3$ could be considered and, if $p_1>0.5$ is thought to be too restrictive, setting $p_1$ as simply the highest weight should typically be enough to identify the model.

\section{Model Estimation}\label{section estim}

Bayesian estimation for the model presented in the previous section is carried out via MCMC, more specifically, a Gibbs sampling algorithm. In order to facilitate the calculations and to make direct simulation from the full conditional distributions in the Gibbs sampling feasible, we introduce three sets of auxiliary variables. The first set was originally proposed by \citet{albert92} for the 2-parameter probit IRT model. This work was later extended to the 3-parameter model by \citet{glas01} to circumvent the intractability introduced by the guessing parameter.

We propose an alternative to the algorithm of \citet{glas01} by also using the auxiliary variable from \citet{albert92} but with a different auxiliary variable to deal with the guessing parameter. This paper will not attempt a systematic comparison between those algorithms. Nevertheless, the dependence structure of the auxiliary variables proposed in our algorithm suggests that a less autocorrelated chain (meaning faster convergence) is obtained when compared to that from \citet{glas01}. Moreover, we also consider lager blocks for the Gibbs sampler, which also induces a less autocorrelated chain.
We define the following auxiliary variables.
\begin{equation}\label{av1}
\ds Z_{ij} \sim Ber(c_i),\;\;\;\ds (X_{ij}|Z_{ij}) \sim N(a_i\theta_j-b_i,1)I_{(Z_{ij}=0)}+\delta_0 I_{(Z_{ij}=1)},
\end{equation}
where $I$ is the indicator function, $Ber$ denotes the Bernoulli distribution and $\delta_0$ is a point-mass at 0 i.e, $P(X_{ij}=0|Z_{ij}=1)=1$.

The auxiliary variables $(Z,X)$ are introduced in the model as follows.
\begin{equation}\label{av2}
Y_{ij}=\left\{
\begin{array}{ll}
\ 1,\;\;\; \mbox{if} \;\; (Z_{ij}=1) \;\; \mbox{or} \;\; (Z_{ij}=0,\;X_{ij}\geq0) \\
\ 0,\;\;\; \mbox{if} \;\; (Z_{ij}=0,\;X_{ij}<0)
\end{array}
\right.
\end{equation}
It is straightforward to check that this model is equivalent to the model in (\ref{m1}) by simply marginalising w.r.t. $(Z,X)$.

Our augmented model (\ref{av1})-(\ref{av2}) makes it feasible to devise a Gibbs sampling for the 3-parameter probit model in which direct simulation from all the full conditional distributions is possible. In order to extend this algorithm to the mixture model proposed in the previous section, we introduce a third set of auxiliary variables to deal with the mixture distribution for the abilities. We define:
\begin{eqnarray}
\ds \pi(\theta_j | W_j,\mu, \sigma)&=& \prod_{k=1}^{K}\ \left[\frac{1}{\sigma_k} \phi \left(\frac{\theta_j-\mu_k}{\sigma_k}\right)\right]^{W_{jk}}, \label{3} \\
  W_j &\sim& Mult(1;p_1,...,p_K),
\end{eqnarray}
which is equivalent to the formulation in (\ref{2}) - $Mult$ represents the multinomial distribution and $\phi$ is the standard normal p.d.f.

The augmented version of our mixture model defines the distribution of the response variables $Y$ in terms of $\Psi=\{a,b,c,\theta,\mu,\sigma^2,p,X,Z,W\}$. Therefore, inference is performed based on the posterior distribution of $(\Psi|Y)$, which has density

\begin{eqnarray}
\ds \pi(\Psi|Y)&\propto&\left[\displaystyle \prod_{i=1}^{I}\prod_{j=1}^{J} \pi(Y_{ij}|X_{ij},Z_{ij})\pi(X_{ij}|Z_{ij},a_i,b_i,\theta_j)\pi(Z_{ij}|c_i)\right]  \nonumber \\
&\times&\left[\displaystyle \prod_{i=1}^{I}\pi(a_i)\pi(b_i)\pi(c_i)\right] \left[\displaystyle \prod_{j=1}^{J}\pi(\theta_j|W_j,\mu,S\sigma) \pi(W_j|p)\right]  \\
&\times&\pi(p) \left[\displaystyle \prod_{k=2}^{K}\pi(\mu_k,\sigma_{k}^2)\right]. \nonumber
\end{eqnarray}
from which we sample through the following MCMC algorithm.

The blocking scheme of our Gibbs sampling algorithm is
\begin{equation}\label{bloc}
\ds\Psi_1=(X,Z),\;\; \Psi_2=(\theta,W),\;\; \Psi_3=(a,b),\;\; \Psi_4=c,\;\; \Psi_5=(\mu,\sigma^2),\;\; \Psi_6=p.
\end{equation}
This scheme takes advantage of the conditional independence among several of model's components, which makes it straightforward to sample from the high dimensional blocks defined in (\ref{bloc}). Such high dimensionality leads to an efficient MCMC algorithm. The full conditional distributions are presented in Appendix A.

Finally, note that the proposed MCMC algorithm is highly parallelisable due to the conditional independence features. More specifically, the update of each of the blocks $\Psi_1$, $\Psi_2$, $\Psi_3$ and $\Psi_4$ can be parallelised since the components inside each of these blocks are conditionally independent for distinct $i$ or $j$. The same works for $\Psi_5$ for each $k$, but with much less impact on the computational cost as $K$ is small.

\section{Simulated examples}\label{section sim}

We present three simulated examples to investigate the efficiency of the methodology proposed in this paper. Each of them consider a mixture distribution for the abilities with glaring multimodality, heavy-tails and skewness, respectively. We also compare the results of the mixture model to the ones from the traditional model (with only one normal) to investigate the impact of model misspecification. In this case, the parameters of the abilities' distribution are fixed to be the mean and variance estimated by mixture model in order to avoid confusion due to different scales when comparing the results. The relative computational time between the mixture model and the normal one is around 1.4.

We consider the following three mixtures to simulate the true value of the abilities:
\begin{equation}\label{mixsim}
0.8N(0,1)+0.2N(2.5,0.5^2),\;\;\;0.7N(0,1)+0.3N(0.5,12^2),\;\;\;0.7N(0,1)+0.3N(1.5,1.8^2). \nonumber
\end{equation}

All three data sets consider 5000 individuals and 50 items and the data is simulated from the 3-parameter probit model. The following prior distributions are adopted: $(\mu_2,\sigma_2^2)\sim NIG(0,100,0.001,0.001)$, $p\sim Beta(2,1)_{I(p_1>0.5)}$, $a_i\sim N(1,3^2)_{I(a_i>0)}$, $b_i\sim N(0,10^2)$ and $c_i\sim Beta(4,12)$. The MCMC chains run for 100 thousand iterations with a burn-in of 50 thousand. Standard diagnostics strongly suggest convergence of the algorithm. The standard raw scores are used as initial values for the abilities. The algorithm is implemented in Ox language \citep{oxdoornik}.\footnote{The original Ox code as well as a BUGS code are available upon request to the authors.}

Results are presented in Table \ref{t1} and Figure \ref{f1}. All the parameters are satisfactorily recovered and the distribution of the abilities is well estimated. Some extra plots regarding the convergence of the chain and the estimation of the abilities and item parameters are presented in Appendix B.

{\renewcommand{\baselinestretch}{1}
\begin{table}[h]\centering
	
	\scriptsize{\begin{tabular}[h]{c|cc|cc}
			
			\hline \\
			&  True mean &  True variance  &  Estimated mean & Estimated variance \\
			\hline\hline \\
			\textbf{\normalsize{Study 1}}		&  \normalsize{0.494}     &   \normalsize{ 1.878} &  \normalsize{0.593} $\scriptscriptstyle(0.511,0.688)$ & \normalsize{2.016}  $\scriptscriptstyle(1.831,2.246)$   \\
			\hline \\
			\textbf{\normalsize{Study 2}}	&   \normalsize{0.110}      &    \normalsize{4.332} & \normalsize{0.011} $\scriptscriptstyle(-0.085,0.105)$& \normalsize{4.215} $\scriptscriptstyle(3.479,4.804)$ \\
			\hline \\
			\textbf{\normalsize{Study 3}}		& \normalsize{0.433}      &  \normalsize{2.119} &  \normalsize{0.341 $\scriptscriptstyle(0.178,0.485)$} & \normalsize{1.984 $\scriptscriptstyle(1.651,2.319)$}   \\
			\hline
		\end{tabular}}
		\caption{True and estimated (posterior mean and 95\% credibility interval) values of the mean and variance of the abilities.}\label{t1}
	\end{table}			
	
For each of the three studies, we compute the \textit{Root Mean Squared Error} (RMSE) - $\sqrt{\ds \sum_{j=1}^{J}(\hat{\theta}_j-\theta_j)^2/J}$, where $\hat{\theta}_j$ is the posterior mean of $\theta_j$. In study 1, the  RMSE was 0.283 and 0.329 for the mixture model and for the traditional normal model, respectively. In study 2, those values were 0.467 and 0.669 and, for study 3, 0.355 and 0.437.

In a real data context, model comparison may be performed through usual (Bayesian) model comparison criteria, for example, DIC and Bayes factor, or by directly comparing the estimated mixture distribution for the abilities to a normal one. The greater the difference is between the two distributions, the higher is the evidence in favor of the mixture model. This could be specifically done by analysing the posterior weight of the first mixture component - smaller values of $p_1$ (away from 1) favor the mixture model.
	
{\renewcommand{\baselinestretch}{1}	
	\begin{figure}[h]
		\centering
		\includegraphics[scale=0.28]{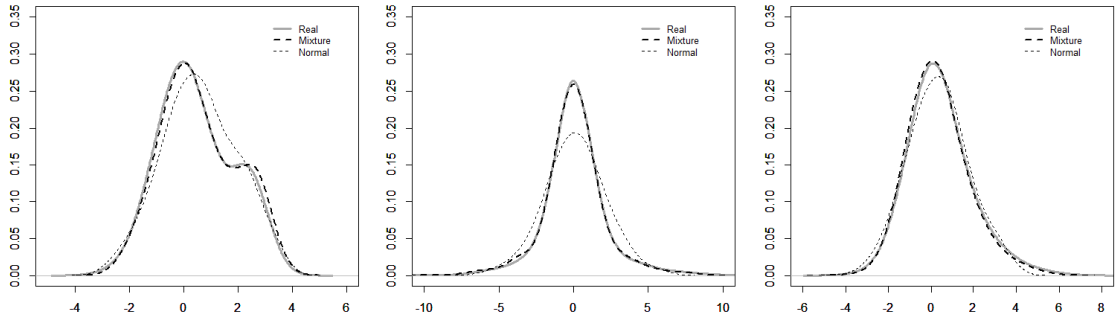}
		\caption{Empirical density of the true and estimated (posterior mean) abilities for the mixture and traditional models.}\label{f1}
	\end{figure}}	
	
We also fit the mixture model to data set generated when the abilities follow a single normal distribution. Results, reported in Appendix B - Figure \ref{apdx1}, show that the abilities are well recovered, despite the model overfit.

The results from this section illustrate the flexibility of the proposed mixture model to consider interesting features (multimodality, heavy tails and skewness) that are not present in the traditional normal model. Furthermore, the disparity between the results of the mixture and normal models (even for a sample of 5000 individuals) highlights the importance of considering the former when the ``true" distribution of the abilities is significantly different from a normal distribution. The results also indicate that the proposed MCMC algorithm is efficient to perform inference. Finally, the proposed model also provides a good fit when the true abilities' distribution is normal. This is quite reasonable as the models are nested.

\section{Real data analysis}

\subsection{ENEM}

We fit our mixture model to a data set from the \emph{High School National Exam} (ENEM) from Brazil. The exam is annually applied to high school students and is organised by the \emph{National Institute of Educational Studies and Researches Anísio Teixeira} (INEP) in the Ministry of Education (MEC). It aims to assess the abilities of the students who are concluding or have already concluded high school in the previous years. The exam is also used in the admission processes in many universities in the country. ENEM is composed of four exams: Humanities, Natural science, Languages and Maths. We consider data from the Maths exam from 2010 consisting of a random sample of 52210 students from São Paulo state. The exam is has 45 items. Figure \ref{srs} shows the histogram of the standard raw scores, which suggests the presence of a positive skewness in the abilities' distribution.
{\renewcommand{\baselinestretch}{1}
\begin{figure}[!h]
	\centering 	\includegraphics[scale=0.45]{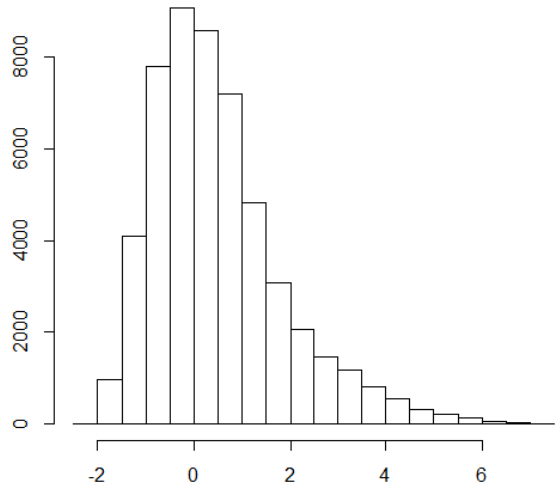}
	\caption{Histogram of the standard raw scores in the Maths exam from ENEM.}\label{srs}
\end{figure}

\begin{table}[!h]\centering
	
	\scriptsize{\begin{tabular}[!h]{ c|cc}
			\hline \\
			&  \textbf{\normalsize{Valor Estimado}}  &      \\  \hline\hline \\
			\normalsize{$\mu_2$}	& \normalsize{1.890}  $\scriptscriptstyle(1.736,2.057)$ &        \\
			\hline \\
			\normalsize{$\sigma_2^2$} 	&   \normalsize{ 2.766}  $\scriptscriptstyle(2.425,3.027)$      &    \\
			\hline \\
			\normalsize{$p_1$}		& \normalsize{0.744}  $\scriptscriptstyle(0.217,0.285)$     &      \\
			\hline
		\end{tabular}}
		\caption{Estimated (posterior mean and 95\% credibility interval) values of the parameters of the abilities' distribution for the ENEM data set.}\label{t3}
	\end{table}

	\begin{figure}[!h]
		\centering 	\includegraphics[scale=0.55]{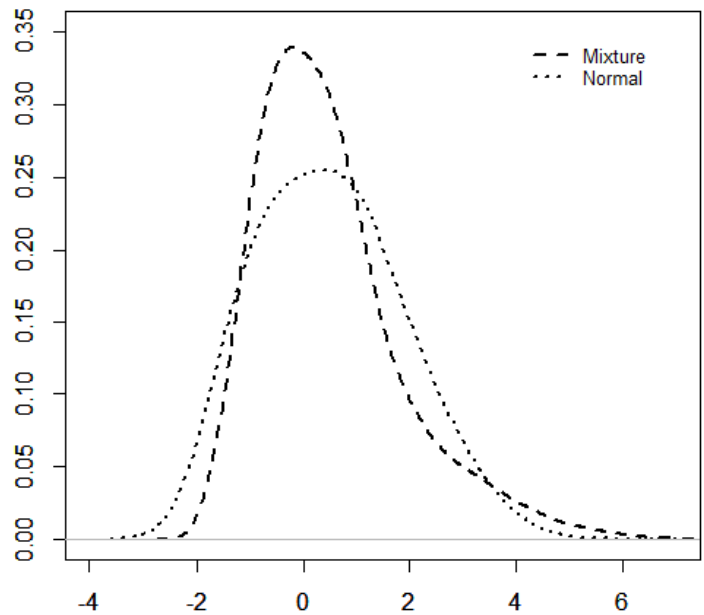}
		\caption{Empirical density of the estimated abilities for the mixture and normal models for the ENEM data set.}\label{f2}
	\end{figure}

The MCMC algorithm runs for 70 thousand iterations with a burn-in of 30 thousand. We adopt the following prior distributions: $(\mu_2,\sigma_2^2)\sim NIG(0,100,0.001,0.001)$, $p\sim Dirichilet(2,1)_{I(p_1>0.5)}$, $a_i \sim N(1,3^2)_{I(a_i>0)}$, $b_i \sim N(0,10^2)$ and $c_i \sim Beta(1,9)_{I(0<c_i<0.15)}$. The restriction on parameters $c_i$ aims at favoring the convergence of the MCMC and is considered to be a reasonable assumption.

We also fit the standard IRT model with no mixture and fix the mean and variance of the abilities to be the ones estimated in the mixture model to eliminate scale interference when comparing the results. Results are presented in Table \ref{t3} and Figure \ref{f2}. The RMSD (\textit{Root Mean Squared Difference}) between the estimated abilities from the mixture model and normal model is 0.28.

\subsection{PISA}

We also analyse some data sets from the PISA exam 2012. We consider the maths test and perform a separate analysis for four countries: Great Britain, South Korea, Poland and United States. The data sets are composed by 12635, 5031, 4596 and 4947 students, respectively. The test has 109 items and each student answers 36 of them. Figure \ref{fpisa} presents the estimated distribution of the abilities and Table \ref{tpisa} some posterior statistics of the mixture parameters. The estimated distributions for GBR and KOR are negatively asymmetric and significantly different from a normal.

Naturally, direct comparisons among different countries cannot be made given that the scale was set within country. We can however analyse the relative discrepancy between the two mixture components across countries to identify different behaviors with respective to features like skewness, modality and heavy tails.

	\begin{figure}[!h]
		\centering 	\includegraphics[scale=0.22]{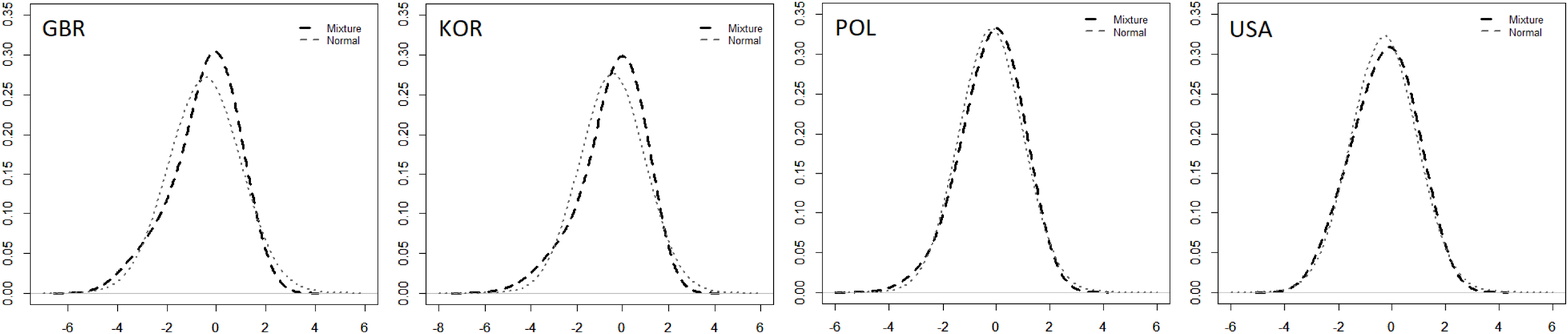}
		\caption{Empirical density of the estimated abilities for the PISA data sets and normal curve with the same mean and variance as the mixture. }\label{fpisa}
	\end{figure}

\begin{table}[!h]\centering
	
	\scriptsize{\begin{tabular}[!h]{c|cccc}
			
			\hline \\
			&  \textbf{\normalsize{$\mu_{2}$}} &  \textbf{\normalsize{ $\sigma^2_{2}$}}  &  \textbf{\normalsize{$p_{1}$}} & \textbf{\normalsize{$p_{2}$}} \\
			\hline\hline \\
			\textbf{\normalsize{GBR}}		
			& \normalsize{-2.495} $\scriptscriptstyle(-3.004,-2.012)$  &  \normalsize{2.171} $\scriptscriptstyle(1.128,3.020)$  &  \normalsize{ 0.814} $\scriptscriptstyle(0.751,0.875)$ & \normalsize{0.186} $\scriptscriptstyle(0.125,0.249)$   \\
			\hline \\
			\textbf{\normalsize{KOR}}		
			& \normalsize{-2.337} $\scriptscriptstyle(-2.916,-1.809)$       &  \normalsize{2.378} $\scriptscriptstyle(1.335,3.471)$  &  \normalsize{0.813} $\scriptscriptstyle(0.732,0.889)$ & \normalsize{0.187} $\scriptscriptstyle(0.111,0.268)$   \\
			\hline \\
			\textbf{\normalsize{POL}}		
			& \normalsize{-2.456} $\scriptscriptstyle(-4.312,-1.347)$ &  \normalsize{1.682} $\scriptscriptstyle(0.137,3.875)$  &  \normalsize{0.916} $\scriptscriptstyle(0.805,0.984)$ & \normalsize{0.084} $\scriptscriptstyle(0.016,0.195)$   \\
			\hline\\
			\textbf{\normalsize{USA}}		
			& \normalsize{-2.246} $\scriptscriptstyle(-2.787,-1.518)$ &  \normalsize{ 0.791} $\scriptscriptstyle(0.121,1.866)$  &  \normalsize{0.869} $\scriptscriptstyle(0.733,0.942)$ & \normalsize{0.131} $\scriptscriptstyle(0.058,0.267)$   \\
			\hline
		\end{tabular}}\caption{Estimated (posterior mean and 95\% credibility interval) values of the parameters of the abilities' distribution for the PISA data sets.}\label{tpisa}
	\end{table}	}

\section{Conclusions}

This paper proposes a generalised approach for the abilities' distribution in dichotomous IRT models. The most general version considers the 3PNO model in which the abilities are assumed to follow a mixture of normal distributions. The methodology is general enough to consider flexible structures for the abilities' distribution, such as multimodality, heavy tails and skewness, without presenting model identifiability problems. That distribution is estimated along with the abilities and item parameters.

Motivated by the need of developing an efficient inference methodology, an MCMC algorithm is proposed to sample from the posterior distribution of all the unknown components of the model. The algorithm is also an efficient alternative to perform inference for the traditional 3PNO model.

The flexibility of the proposed model as well as the efficiency of the MCMC algorithm are illustrated in simulated examples. By comparing the results to the ones from the traditional model, we highlight the importance of considering a more general approach.
Finally, the proposed model is applied to real data examples and unveils non-normality behaviours.

The mixture approach for the abilities proposed here can be, sometimes straightforwardly, considered in other families of IRT models. One interesting example would be the 4PNO model, which considers an upper asymptote \citep[see][]{culpepper1,culpepper2}.

\section*{Appendix A - MCMC details}

We now present the full conditional distributions of each block from the Gibbs sampling algorithm proposed in Section \ref{section estim}. We assume that all the $J$ individuals answer all the $I$ items. Adaptations are straightforward if that is not the case by simply ignoring the respective likelihood terms.

The pairs $(Z_{ij},X_{ij})$ are conditionally independent for all $i$ and $j$, which means that we can sample the vector $(Z,X)$ by sampling each pair $(Z_{ij},X_{ij})$ individually. Moreover,
\begin{equation*}
\ds \pi(Z_{ij},X_{ij}|.)= \left\{
\begin{array}{ll}
\ds\frac{\phi(x_{ij}-m_{ij})}{\Phi(-m_{ij})} I_{(Z_{ij}=0)} I_{(X_{ij}<0)},\ \ \ if\ Y_{ij}=0 \\
\ds w_{ij}I_{(Z_{ij}=1)}I_{(X_{ij}=0)}+(1-w_{ij})\frac{\phi(x_{ij}-m_{ij})}{\Phi(m_{ij})}I_{(Z_{ij}=0)}I_{(X_{ij}>0)},\ \ \ if\  \ Y_{ij}=1,
\end{array}
\right. \label{6}
\end{equation*}
where $\ds m_{ij}=a_i\theta_j-b_i$ and $\ds w_{ij}=\frac{c_i}{c_i+(1-c_i)\Phi(m_{ij})}$. This means that, if $Y_{ij}=0$, $Z_{ij}$ is a point-mass at zero and $X_{ij}$ is a $N(m_{ij},1)$ truncated to be less than 0. On the other hand, if $Y_{ij}=1$, $Z_{ij}$ is a $Ber(w_{ij})$ and $X_{ij}$ is a point-mass at 0 if $Z_{ij}=1$ and is $N(m_{ij},1)$ truncated to be greater than 0 if $Z_{ij}=0$.\\

All the $c_i$ parameter are conditionally independent with
\begin{equation*}
\ds (c_i|\cdot) \sim Beta \left(\displaystyle \sum_{j=1}^{J}Z_{ij}+\alpha_{c},\ J-\displaystyle \sum_{j=1}^{J}Z_{ij}+\beta_{c}\right) \label{7}
\end{equation*}

The pairs $(a_i,b_i)$ are also conditionally independent with
\begin{equation*}\label{8}
\ds (a_i,b_i|\cdot) \sim \ N_2(\mu_i,\Sigma_i)I_{a_i>0},\;\;\mbox{for } \mu_i=\left[\begin{array}{r}{\mu_a}^*,\;\;{\mu_b}^*\end{array}\right],\;\; \Sigma_i=\left[\begin{array}{rr} {\sigma_a^2}^*&\gamma\\\gamma&{\sigma_b^2}^*\end{array}\right],
\end{equation*}\\
where $\ds{\sigma_a^2}^*=\frac{\sigma_a^2}{\left(\sigma_a^2\displaystyle \sum_{j=1}^{L_i} \theta_j^2+1\right) \left(1-\gamma^2\right)}$, $\ds{\sigma_b^2}^*=\frac{\sigma_b^2}{\left(L_i\sigma_b^2+1\right) \left(1-\gamma^2\right)}$,\\$\ds\gamma=\frac{\sigma_a\sigma_b\displaystyle \sum_{j=1}^{L_i} \theta_j}{\left[\left(\sigma_a^2\displaystyle \sum_{j=1}^{L_i} \theta_j^2+1\right)\left(L_i\sigma_b^2+1\right)\right]^\frac{1}{2}}$,\\
$\ds{\mu_a}^*={\sigma_a^2}^*(\displaystyle \sum_{j=1}^{L_i} x_{ij}\ \theta_j+\mu_a \sigma_a^{-2}) - {\sigma_a}^*\  {\sigma_b}^*\ \gamma\ (\displaystyle \sum_{j=1}^{L_i} x_{ij} - \mu_b \sigma_b^{-2})$ and\\
$\ds{\mu_b}^*={\sigma_a}^*\  {\sigma_b}^*\ \gamma\ (\displaystyle \sum_{j=1}^{L_i} x_{ij}\ \theta_j+\mu_a \sigma_a^{-2}) - {\sigma_b^2}^* \ (\displaystyle \sum_{j=1}^{L_i} x_{ij} - \mu_b \sigma_b^{-2})$, with $L_i$ referring to the individuals for which $Z_{ij} = 0\}$.
We sample from this distribution via rejection sampling by proposing from the unrestricted distribution and accepting if $a_i>0$.\\

The pairs $(\mu_k,\sigma_{k}^2)$ are also conditionally independent with
\begin{equation*}
\ds (\mu_k,\sigma^2_k|\cdot) \sim NIG \left(m^*; \ \frac{1}{\beta^*}; d^*;  e^*\right),
\end{equation*}
where $\ds m^*=\frac{\displaystyle \sum_{j=1}^{J} W_{jk} \theta_j + m\beta}{\displaystyle \sum_{j=1}^{J} W_{jk} + \beta} $, \ $\ds\beta^*=\displaystyle \sum_{j=1}^{J} W_{jk} + \beta$, \ $\ds d^*=d+ \displaystyle \frac{1}{2}\left(\sum_{j=1}^{J} W_{jk}\right)$, \\$\ds e^*=e + \frac{\displaystyle \sum_{j=1}^{J} W_{jk}\beta}{2\left(\beta +\displaystyle \sum_{j=1}^{J} W_{jk}\right)}\left(m-\bar{\theta}\right)^2 + s/2$, where\\ $\ds s=\displaystyle \sum_{j=1}^{J} W_{jk} \theta_j^2 - \displaystyle \sum_{j=1}^{J} W_{jk}$,\;  $\ds \bar{\theta}=\frac{\displaystyle \sum_{j=1}^{J} W_{jk}\theta_j}{\displaystyle \sum_{j=1}^{J} W_{jk}}$.\\

The full conditional distribution of $p$ is
\begin{equation*}
\ds (p|\cdot) \sim Dirichilet \left(\displaystyle \sum_{j=1}^{J} W_{j1} + \alpha_1 ,\dots,   \displaystyle \sum_{j=1}^{J} W_{jK} + \alpha_K\right)I_{(p_1>0.5)}
\end{equation*}
We sample from this distribution via rejection sampling by proposing from the unrestricted distribution and accepting if $p_1>0.5$.\\

The pairs $(\theta_j,W_j)$ are also conditionally independent with
\begin{equation*}
\ds (\theta_j|W_j,\cdot) \sim N(\mu^{*}_k,\sigma^{*}_k),\;\; W_j\sim\pi_{Mult}(1,p^{*}_1, \dots,p^{*}_K),
\end{equation*}
where $\ds{p_k}^{*}=\frac{\alpha^{*}_k}{\displaystyle \sum_{k=1}^{K} \alpha^{*}_k}$,\  $\ds\mu^{*}_k=\frac{\mu_k+\sigma^2_k \displaystyle \sum_{i=1}^{L_j}a_i(x_{ij}-b_i)}{1+\sigma^2_k \displaystyle \sum_{i=1}^{L_j} {a_i}^2}$,  $\ds\sigma^{2*}_k=\frac{\sigma^2_k}{1+\sigma^2_k \displaystyle \sum_{i=1}^{L_j} {a_i}^2}$ and

\begin{eqnarray*}
\ds\alpha^{*}_k &=& p_k \ \left(1+\sigma^2_k \displaystyle \sum_{i=1}^{L_j} {a_i}^2\right)^{-\frac{1}{2}} \exp\left\{ -\frac{1}{2}  \left[ \frac{\mu^2_k}{\sigma^2_k}    - \frac{\left(\mu_k+\sigma^2_k \displaystyle \sum_{i=1}^{L_j}a_i(x_{ij}-b_i) \right)^2}{\sigma^2_k\left(1+\sigma^2_k \displaystyle \sum_{i=1}^{L_j} {a_i}^2\right)}      \right]    \right\}, \nonumber
\end{eqnarray*}
where $L_j$ refers to the items for which $Z_{ij}=0$.

\newpage

\section*{Appendix B - Further results from the simulations}

{\renewcommand{\baselinestretch}{1}

	\begin{figure}[!h]
		\centering
		\includegraphics[scale=0.27]{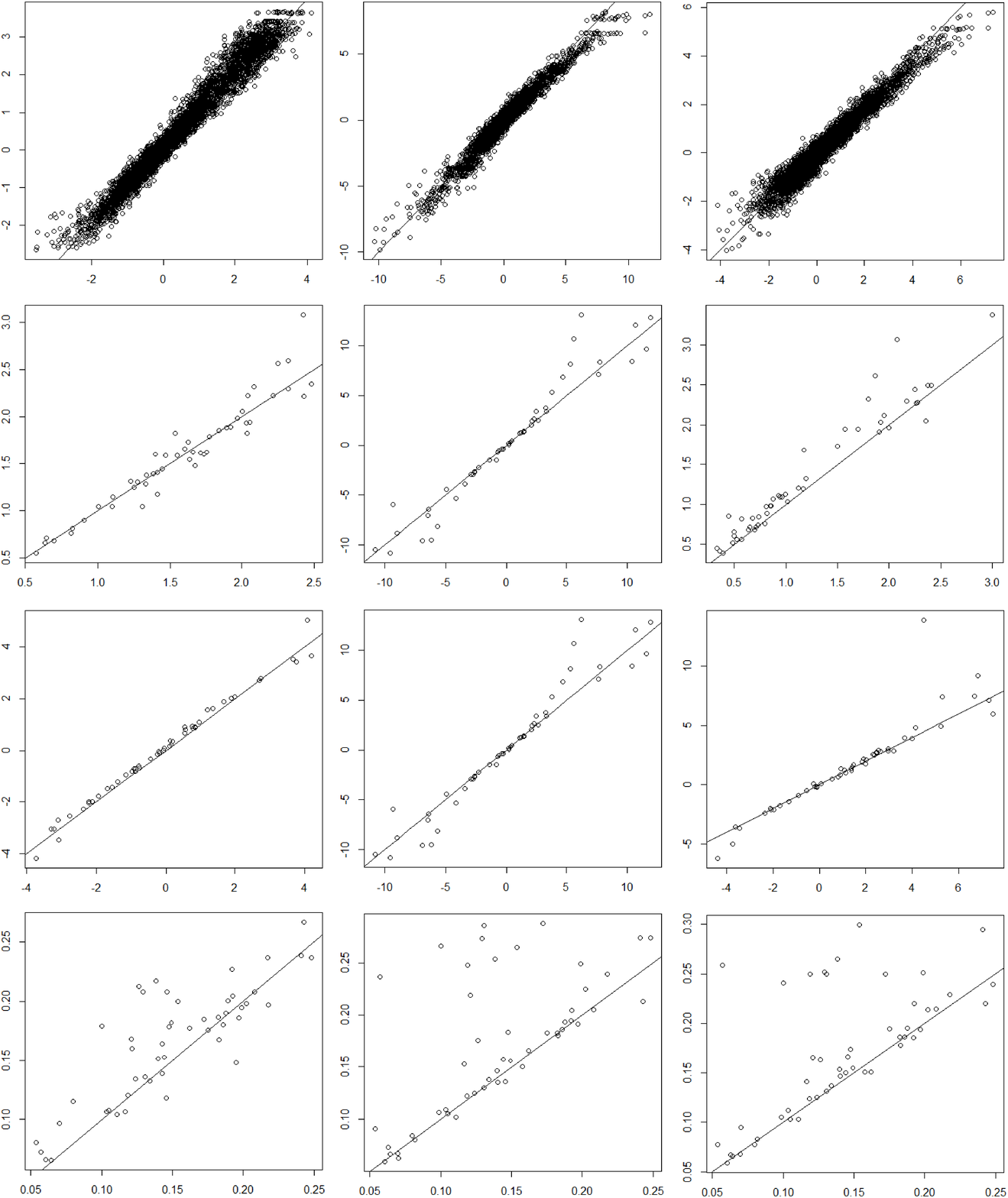}
		\caption{True (x-axis) \textit{versus} estimated (posterior mean) value of $\theta$, $a$, $b$, $c$ for the three studies (one per column).}
	\end{figure}

	\begin{figure}[!h]
		\centering
		\includegraphics[scale=0.5]{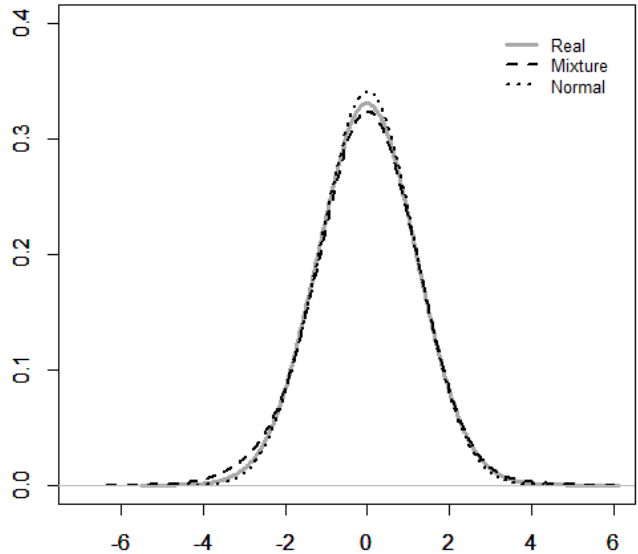}
		\caption{True and estimated abilities' distribution when data is generated from the standard model but the mixture model is fit. The true distribution is Normal with the mean and variance of the estimated mixture (to avoid confusion due to scale).}\label{apdx1}
	\end{figure}

	\begin{figure}[!h]
		\centering
		\includegraphics[scale=0.70]{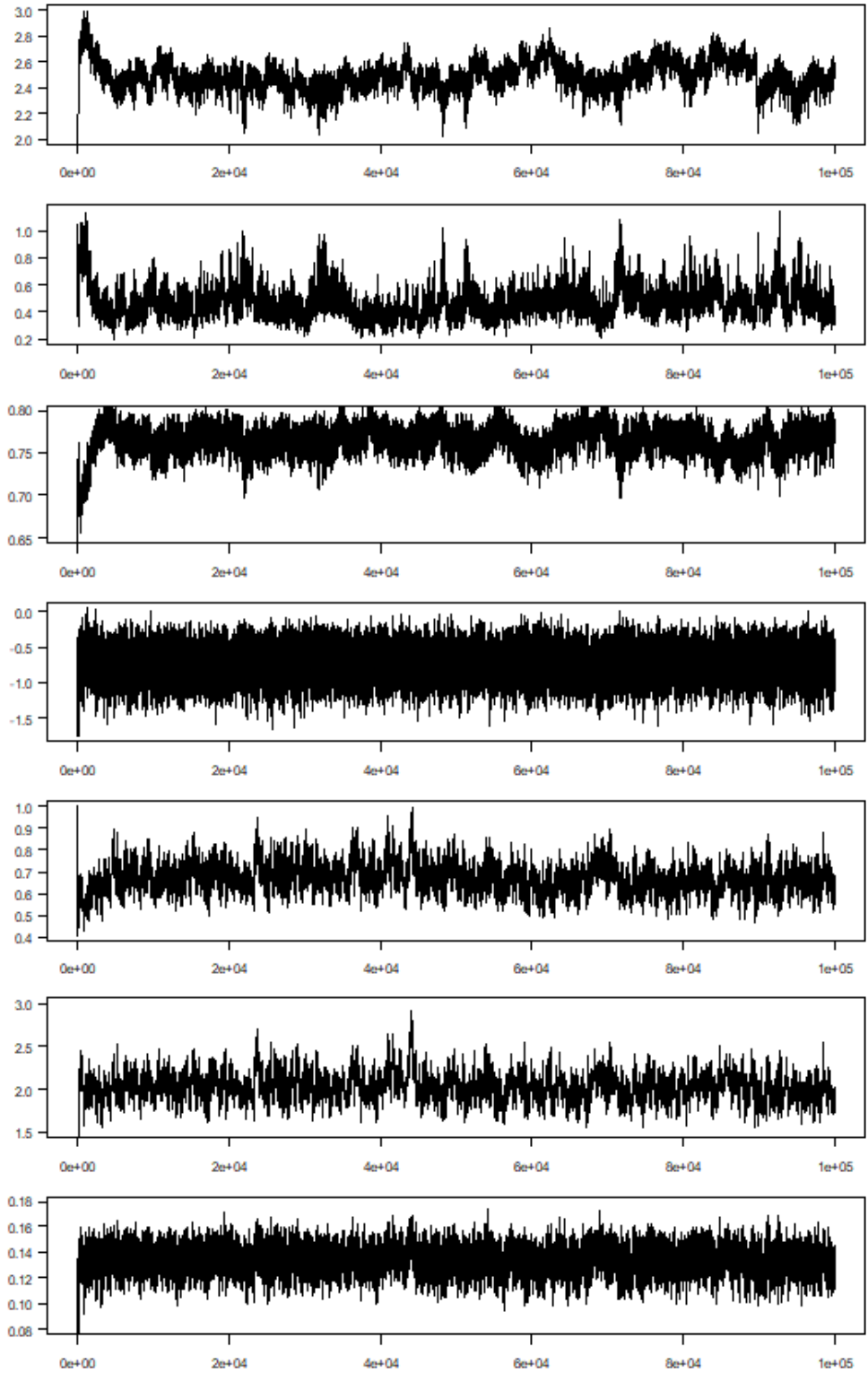}
		\caption{Trace plots for $\mu^2$, $\sigma_{2}^2$, $p_1$, $\theta_{100}$, $a_{2}$, $b_{2}$, $c_{2}$ in study 1.}
	\end{figure}
}

\vspace{3cm}

\bibliographystyle{apalike}
\bibliography{refs}

\end{document}